\documentclass[aps,prl,preprint,showpacs,floatfix]{revtex4}
\usepackage{graphicx}

\begin{document}

\title{Is Correlation Important in Explaining the Charge Transport in Disordered Molecular Solids ?} 

\author{Ajit Kumar Mahapatro and Subhasis Ghosh}

\affiliation{School of Physical Sciences, Jawaharlal Nehru University,
New Delhi 110067, India}

\begin{abstract}
We have examined the electrical transport in disordered molecular solids. It has been found that mobility is a  function of electric field and temperature due to hopping conduction. Several theoretical models for charge transport in disordered solids have been debated over the role of spatial and energetic correlation in these systems and  such correlations have been recently shown to explain the universal electric field dependence of mobility. We have compared and evaluated the applicability of different theoretically proposed models using very simple experimental results and based on our extensive analysis, we have found that correlation is important to explain the electrical transport in these systems.
\end{abstract}


\maketitle

Recently, considerable debate\cite{dhd,snn,pep} is going on regarding the charge carrier transport in disordered organic molecular solids(DOMS) because of two reasons: first, their application in display devices and second, the fundamental understanding of charge transport in these materials. There are several distinguishing features of these solids, (i) they are composed of organic molecules, held together loosely by   weak van der Waals type intermolecular coupling while the intramolecular coupling is strong; (ii) the absence of long range order in these disordered materials lead to the localization of the electronic wave function and the formation of a broad Gaussian density of states(GDOS) and  the most important one is (iii) the carrier mobility $\mu$ exhibits a nearly universal Poole-Frenkel(PF) behavior\cite{lbs,dhd1}
\begin{equation}
\mu(E)=\mu(0) \exp\left(\gamma \sqrt{E}\right)
\end{equation}
\noindent where  $\mu(0)$ and $\gamma$ are temperature dependent quantities, known as the zero field mobility and the field activation of the mobility, respectively.  In DOMS,  hopping among the molecular sites having comparable energies, describes the transport of charge carriers through the GDOS of highest occupied molecular orbital(HOMO) and/or lowest unoccupied molecular orbital(LUMO).

There are several models to explain the universal feature of carrier mobility $\mu$(Eq.1) observed in wide class of organic solids. Gill\cite{wdg} first attempted to describe the experimental data with the following empirical form of zero field mobility $\mu(0)$ and field activation of mobility $\gamma$, $\mu(0)=\mu_0 exp[-\Delta/(k_BT)]$ and $\gamma=B [1/(k_BT)-1/(k_BT_0)]$, where $\mu_0$ is the zero field temperature independent mobility, 
$\Delta$ is the zero-field activation energy, and $B$ and $T_0$ are parameters of the model.
 There are several conceptual problems with this phenomenological description and the  most important are (i) the empirical Gill's formula lacks theoretical justification and the significance of different parameters($\Delta$, $B$ and $T_0$) are not well understood in the context of the physical properties of the system and (ii) the empirical formula for $\gamma$ predicts the negative field dependence for $T\geq T_0$, which deviates from the universal feature(PF) of carrier transport in these materials.  It is now recognized that the spatial fluctuations in the potential energy of the charge carriers in DOMS results the PF behavior of carrier mobility\cite{dhd,snn,pep}. There are two competing models, debating over the role of correlation on the spatial fluctuations in these systems.
Bassler and co-workers\cite{hb} proposed   the uncorrelated Gaussian disorder model(UGDM) and provided support to PF behavior of mobility using Monte Carlo simulation. The UGDM describes the carrier transport as a biased random walk among the dopant molecules with Gaussian-distributed random site energies. UGDM lead to following form of $\mu(0)$ and $\gamma$
\begin{equation}
\mu(0)=\mu_0 \exp \left \{-\left(\frac{2}{3} \hat{\sigma} \right )^2 \right \}, 
 \gamma=C(\hat{\sigma}^2-\Sigma)
\end{equation}
\noindent where $\hat{\sigma}=\sigma/kT$, $\sigma$ is the width of the Gaussian distribution of the hopping sites energies, and $C$ and $\Sigma$ are parameters of the model. In this case, $\mu(0)$ and $\gamma$ have  $1/T^2$ dependence, instead of $1/T$ dependence as in case of Gill's empirical formula. Although UGDM explains\cite{hb} some features of experimental data and provides support for PF behavior of carrier mobility, several discrepancies emerge with uncorrelated description of Gaussian disorder model, which will be discussed later.  The most important criticism\cite{lbs1} against UGDM is its' inability to reproduce the PF behavior over wider range of electric field. 
Garstein and Conwell\cite{yng} first showed that a spatially correlated potential is required for the description of PF behavior of mobility for wider range of electric field.  Dunlop and co-workers\cite{dhd,snn} have shown analytically in 1D and numerically in 3D that the interaction of charge carriers with permanent dipoles located on either dopant or host molecules give rise to PF behavior of mobility. Essentially, this correlated Gaussian disorder model(CGDM) is based on long-range correlation between charge carriers and the molecular electric dipole. They have shown that this interaction gives rise to random potential energy landscape with long-range spatial correlations $<U(0)U(r)>\sim\sigma^2a/r$, where a is the minimal charge-dipole separation or the lattice constant of the molecular solid. CGDM gives the following form of $\mu(0)$ and $\gamma$
\begin{equation}
\mu(0)=\mu_0 \exp \left \{-\left(\frac{3}{5}\hat{\sigma}\right)^2 \right \}, \gamma=A\left ( \hat{\sigma}^{3/2}-\Gamma\right ) \left (\frac{ea}{\sigma}\right )^{1/2}
\end{equation}
\noindent where $A$ and $\Gamma$ are  parameters of the model. Here, $\mu(0)$ has a $1/T^2$ dependence(same as in UGDM) but $\gamma$ has a $1/T^{3/2}$ dependence. In this letter, we present the measurements on charge carrier mobility $\mu$ and its field activation $\gamma$ as a function of temperature and electric field. The experimental results are analyzed within UGDM and CGDM. It has been found that correlated transport model CGDM successfully explains the experimental data emphasizing the important role of correlation in explaining the transport mechanism in disordered organic solids.

We have chosen two different metal-phthalocyanines(MePc), Copper-phthalocyanine(CuPc) and Zinc-phthalocyanine(ZnPc) for our investigations because of their high chemical stability, ease to prepare the devices and reproducibility of experimental data, which is a major problem with most other organic solids due to their degradation with time. Details about the MePc-based single layer devices are given in Ref.\cite{akmsg} and Ref.\cite{akmsg1}.
In this letter, we report the experimental investigation on charge carrier  transport in hole only
devices based on metal/MePc/metal
structures. In this case, by properly choosing contacting metals,
current injection and transport due to holes have been
investigated.  Indium tin
oxide(ITO) has work function(4.75eV) very close to the ionization
potential(4.8eV) of MePc\cite{gp}, resulting the ITO/MePc interface as an Ohmic contact.
Fig.1 shows the current-voltage(J-V) characteristics of ITO/CuPc/Al
and ITO/CuPc/Cu at room temperature. The experiment consisted of
two steps. First: the current due to hole injection from ITO was measured by 
supplying it to positive bias 
and second: the current due to hole injection from Cu and Al was measured
by reversing the polarity of the bias voltage(i.e., biasing Al and Cu electrodes positively). It is clear from  Fig.1
that J-V characteristics in case of ITO/MePc/Cu display almost symmetric
behavior in both cases(hole injection either from ITO, or from Cu
electrodes). This is because in both cases there is small energy
barriers of 0.05eV in case of ITO/MePc  and 0.2eV in case of Cu/MePc interface, giving rise to space charge limited(SCL) bulk  current when either ITO or Cu is positively biased. It has been shown\cite{akmsg, psd}, for Schottky energy barrier(SEB) less than about 0.3-0.4eV, the current flow is due to SCL\cite{peb,ajc,pmbw}.
 But, in case of ITO/MePc/Al devices, J-V characteristics displays asymmetric and rectification-like
behavior. Current density increases by almost five order of
magnitude by reversing the polarity of the bias. As discussed before,
when ITO is positively biased, current is due to SCL and when Al
is positively biased, current is due to injection limited and
reduced by several order of magnitude due to the existence of SEB
of 0.6eV at MePc/Al contact. It has been shown by us\cite{akmsg} that this injection limited current $J_{inj}$ from a metal electrode to  disordered organic semiconductor can be described by the relation,  $J_{inj}\propto\psi^2\mu exp(-\phi_B/k_BT)exp(f^{1/2})$, $\psi$ is a slow varying function of E, $\phi_B$ is SEB and $f=e^3E/2\pi\epsilon(k_BT)^2$, proposed by Shen {\sl et. al.}\cite{yshen}. 

Inset of Fig.1 shows the temperature dependent SCL current in  ITO/MePc/Al structure made out of 200nm thick active organic layer of CuPc. At low bias, J-V characteristic follow the slope of 2 in log(J)-log(V) plot, but as the bias increases the slope   increases gradually to higher value. A numerical workout\cite{akmsg1} has been  used by solving  Poission's equation, $dE(x)/dx=ep(x)/\epsilon$ and $J(x)=ep(x)\mu(x)E(x)$ simultaneously,  and assuming PF form(Eq.1) of the carrier mobility. Here, $\epsilon$ is the dielectric constant of MePc and $p(x)$ is the hole density at position x and  the boundary condition is taken  at the Ohmic contact(ITO/MePc interface) with  hole density(at $x=0$) of   $2.5\times 10^{19} cm^{-3}$, which is the density of states of HOMO  in MePc. The simulated results are shown as solid lines in inset of Fig.1 at different temperatures.  
The excellent  agreement of the simulated and the experimental data confirms PF behavior of carrier  mobility and the thermally activated  hopping transport.  Fig.2 shows the temperature dependence of conductivity $\sigma(T)$ for CuPc. Linear dependence of $log(\sigma)$ on $1/T^{1/4}$  validates the  variable range hopping(VRH)\cite{nfm}, $\sigma(T)=\sigma_0 exp\{-(T_1/T)^{1/4}\}$, where $\sigma_0$ is the temperature independent conductivity and $T_1$ is the characteristic temperature. The inset shows the dependence of $\sigma_0$ at different electric field and linear dependence of $log(\sigma_0)$ on $\sqrt{E}$ signifies the PF behavior in charge carrier transport conclusively. Hence the experimental data presented in Fig.1 and Fig.2 establish the carrier transport process, which is thermally activated hopping with field dependent mobility. 

Now, we compare our experimental data with the  uncorrelated(UGDM) and correlated(CGDM) theoretical models proposing the stretched exponential electric field  dependence of $\mu$.
At every temperature,  $\mu(0)$ and  $\gamma$ are obtained from the experimental data and plotted in Fig.3 and Fig.4. Fig.3 shows the temperature dependence of the $\mu(0)$  and  $\gamma$ according to  UGDM and the apparent linear dependence of ln$\mu(0)$ on $1/T^2$ and $\gamma$ on $1/T^2$ can be described by  Eq.(2).
Values  of temperature independent mobility $\mu_0$, width of the energy spread $\sigma$, and the constant $C$ and $\Sigma$ are determined from the  ln$\mu(0)$ vs. $1/T^2$ and $\gamma$ vs. $1/T^2$ plots, and given in Table I.  Main feature of UGDM is the non-Arrehenious behavior of the temperature dependence of mobility, which is the consequence of hopping conduction in GDOS. But, there are several discrepancies in this description, (i) fitting is not as good as in case of CGDM, which will be discussed next; (ii) the intercept of $\gamma$ vs. $1/T^2$(shown in Fig.3) gives positive value resulting negative value of parameter $\Sigma$(given in Table I), but according to UGDM(Eq.2), it should be negative, which is a fundamental problem with UGDM, and (iii) it has been observed that there is no specific trend in the values of $C$ and $\Sigma$, which is a scaling parameter in the model and cannot be linked to the physical parameters of the system.

Fig.4 shows the temperature dependence of $\mu(0)$ and $\gamma$ according to  CGDM, described by Eq.(3).  Values of the parameters $\mu_0$, $\sigma$,  $a$  and $\Gamma$  are determined from the straight line plots of  ln$\mu(0)$ vs. $1/T^2$ and $\gamma$ vs. $1/T^{3/2}$ and given in Table I. Following observations have been made regarding the CGDM description of the experimental data, (i)  linear dependence of ln$\mu(0)$ on $1/T^2$ and $\gamma$ on $1/T^{3/2}$ is evident in Fig.4 and straight line fit is excellent and better than that in UGDM; (ii) for all samples the intercept of $\gamma$ vs. $1/T^{3/2}$ gives negative values, according to CGDM(Eq.3);   (iii) the experimental value of $\Gamma$ are 1.2 in CuPc and 1.4 in ZnPc, which are very close to the predicted value 1.97\cite{snn};  (iv)  the width of the Gaussian distribution $\sigma$ is 100meV in CuPc and   120meV in  Zn-Pc and similar values for $\sigma$ are commonly observed in other molecular solids;    (v) the lattice constant($a$) has been found to be 19$\AA$ in CuPc and 17$\AA$ in ZnPc, are in excellent agreement with the reported\cite{rr} experimental values determined by X-Ray diffraction and (vi)  MePc molecules are symmetric, so the dipole moment $p$ should be zero or very small, but the symmetry of the individual molecule  may be lowered in solid state,  so, in any case $p$ should be very small, which is corroborated by the estimated small value of $p$(=0.3D) for CuPc using the relation $\sigma=2.35ep/\epsilon a^2$, according to CGDM\cite{snn}.    

The UCDM and CGDM share the common feature  regarding the temperature dependence of  $\mu(0)$ due to similar distribution of hopping sites energies, but the temperature dependence of field activation   parameter $\gamma$ decides critically the importance of spatial correlation and decides the mechanism of carrier transport in organic molecular solids. Though CGDM explains the experimental data successfully, there are some limitations with this model.  The positional disorder is not included in the present form of CGDM, but large values of disorder parameters  $\sigma$ and $\Gamma$  in NPPDA\cite{snn}, AlQ\cite{ggm} and  MePc  signify the importance of the positional disorder. In CGDM, it has been shown that long-range interaction between charge carriers and permanent dipole moments of doped molecules in polymers and  host molecules leads to spatial correlation. However, it has been pointed out by Yu {\sl et. al.}\cite{zgy} that the mechanism responsible for PF behavior in different conjugated polymers and molecules cannot be due to only charge-dipole interaction, because  PF behavior has been universally observed in several doped and undoped polymers and molecules with or without permanent dipole moment.  Hence, in addition to charge-dipole interaction there may be another mechanism responsible for spatial correlation, which is a fundamental requirement for PF behavior. Yu {\sl et. al.}\cite{zgy} have  shown using first principle quantum chemical calculation that the thermal fluctuations in the molecular geometry can lead to spatial correlation. It has been shown that the primary restoring force for the thermal fluctuation is steric and intermolecular, which lead to spatially correlated fluctuation in the energies of the localized states. Further experimental investigations are in progress to identify the exact origin of spatial correlation in molecules with and without dipole moments.

In summary, we have studied the mechanism of carrier transport in disordered molecular solids by means of simple  experiments. We have (i) shown that the carriers conducts through hopping; (ii) given evidences for the universal electric field dependence($ln(\mu) \propto \sqrt{E}$) of carrier mobility and (iii) presented a comparison of the transport properties with uncorrelated and correlated disorder model. We have shown  that $E$ and $T$ dependence of experimental data can be fitted and described within the correlated Gaussian disorder model, signifying the important role of both energetic and spatial disorder on the carrier transport in these materials. The origin of correlation has been discussed. We hope present analysis would lead to further experimental and theoretical investigations focusing the exact origin and role of correlation on transport properties of disordered molecular solids.
 
\vspace{0.2in}

\noindent{\sl Table.I. Temperature independent zero field mobility $\mu(0)$, width of energy spread $\sigma$($\sigma_I$ in UGDM and $\sigma_{II}$ in CGDM)  and different scaling parameters $C$, $\Sigma$ in UGDM and $a$, $\Gamma$ in CGDM for CuPc and ZnPc are given.}
\begin{center}
	\begin{tabular}{cccccccc}\hline \hline
 &$\mu_0$ & $\sigma_I$ & $C$ &  & $\sigma_{II}$ & $a$ &  \\
 & in  & in & in & $\Sigma$ & in & in & $\Gamma$ \\
 &$\frac{cm^2}{Vs}$ & meV & ($\frac{cm}{V}$)$^{\frac{1}{2}}$ & & meV  & $\AA$ & \\ \hline	
~~CuPc ~~ & $~~3.0\times 10^{-8}~~$ & ~~95~~ & ~~$ 6.9\times 10^{-4}$~~ & ~~-3.18~~ & ~~100~~ & ~~19~~ & ~~1.4~~ \\	
~~ZnPc ~~& $~~1.7\times 10^{-8}~~$ & ~~110~~ & ~~$ 4.5\times 10^{-4}$~~ & ~~-4.24~~ & ~~120~~ & ~~17~~ & ~~1.2~~ \\			\hline \hline
	\end{tabular}
\end{center}

\newpage

\noindent {\large \bf Figure caption.}

\vspace{0.2in}

\begin{description}

\item[Fig.1.] Room temperature J-V characteristics for single layer hole only CuPc device with thickness of 200nm.  Empty circles represent the data when  ITO is positively biased and empty squares represent the data when ITO is negatively biased.  Inset shows the J-V characteristics of bulk limited current in  ITO/CuPc/Al  at different temperature staring at 320K and then at the interval of 40K. Empty circles show experimental data and solid lines represent the  simulated data. 

\item[Fig.2.] Temperature dependence  of conductivity of CuPc. Empty circles are experimental data and solid line is fit with straight line. Inset shows the electric field dependence of  temperature independent conductivity. Solid line is fit with PF dependence of mobility.

\item[Fig.3.] Temperature dependence of  the field activation parameter $\gamma$ and zero field mobility $\mu(0)$(inset) for CuPc obtained from the theoretical fit to the experimental data,  according to the UGDM(Eq.2).

\item[Fig.4.] Temperature dependence of  the field activation parameter $\gamma$ and zero field mobility $\mu(0)$(inset) for CuPc obtained from the theoretical fit to the experimental data, according to the CGDM(Eq.3).

\end{description}

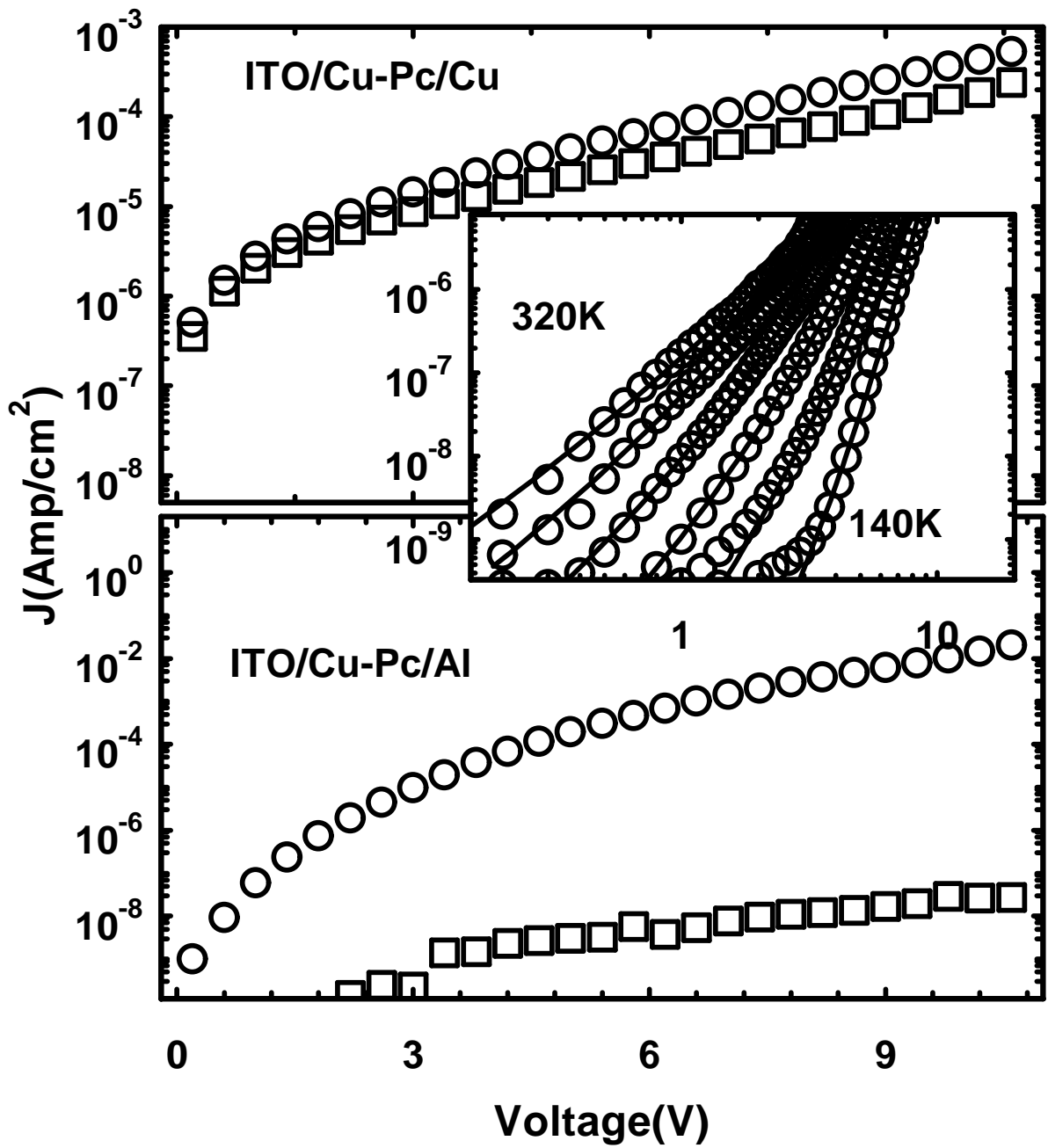
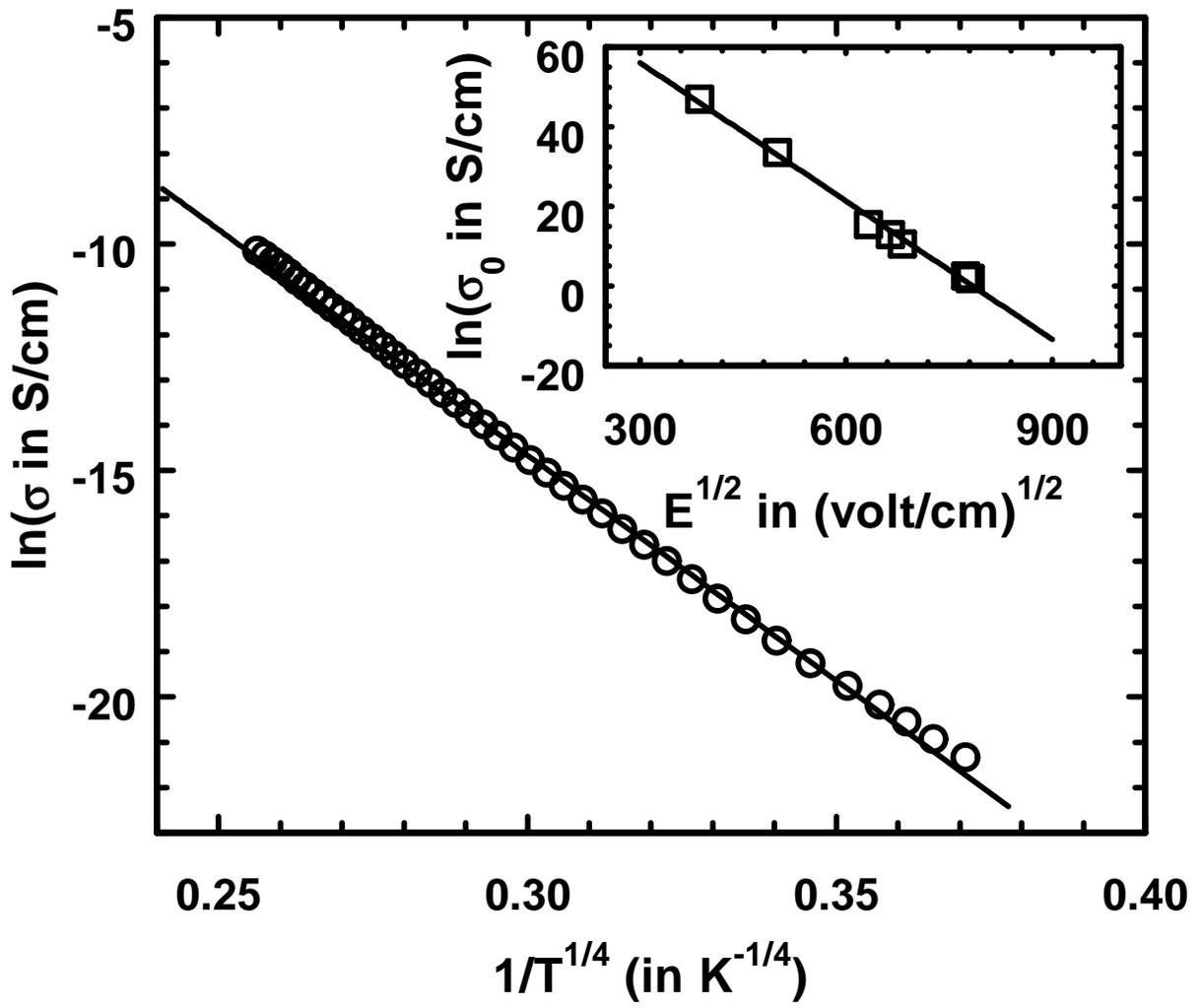
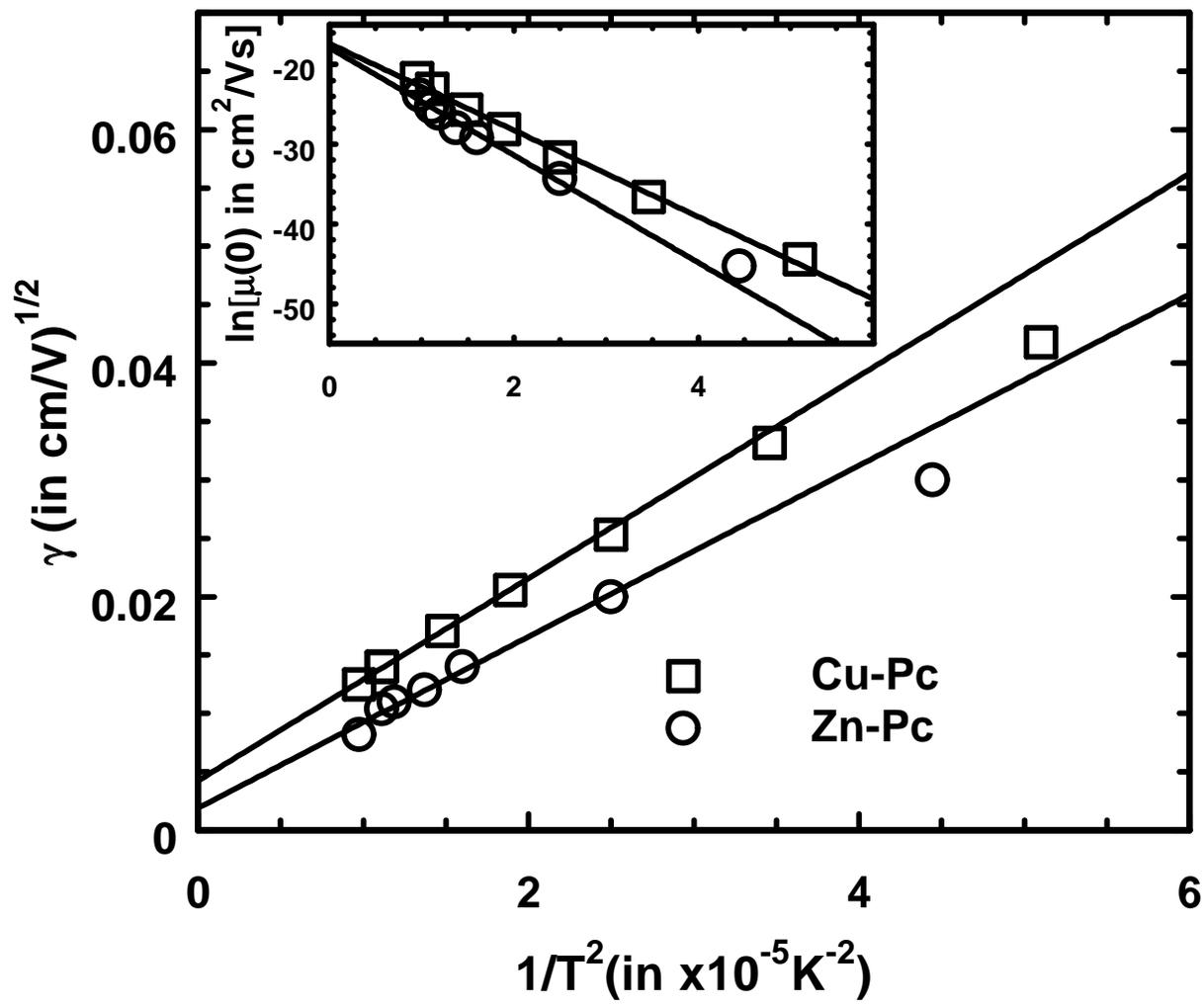
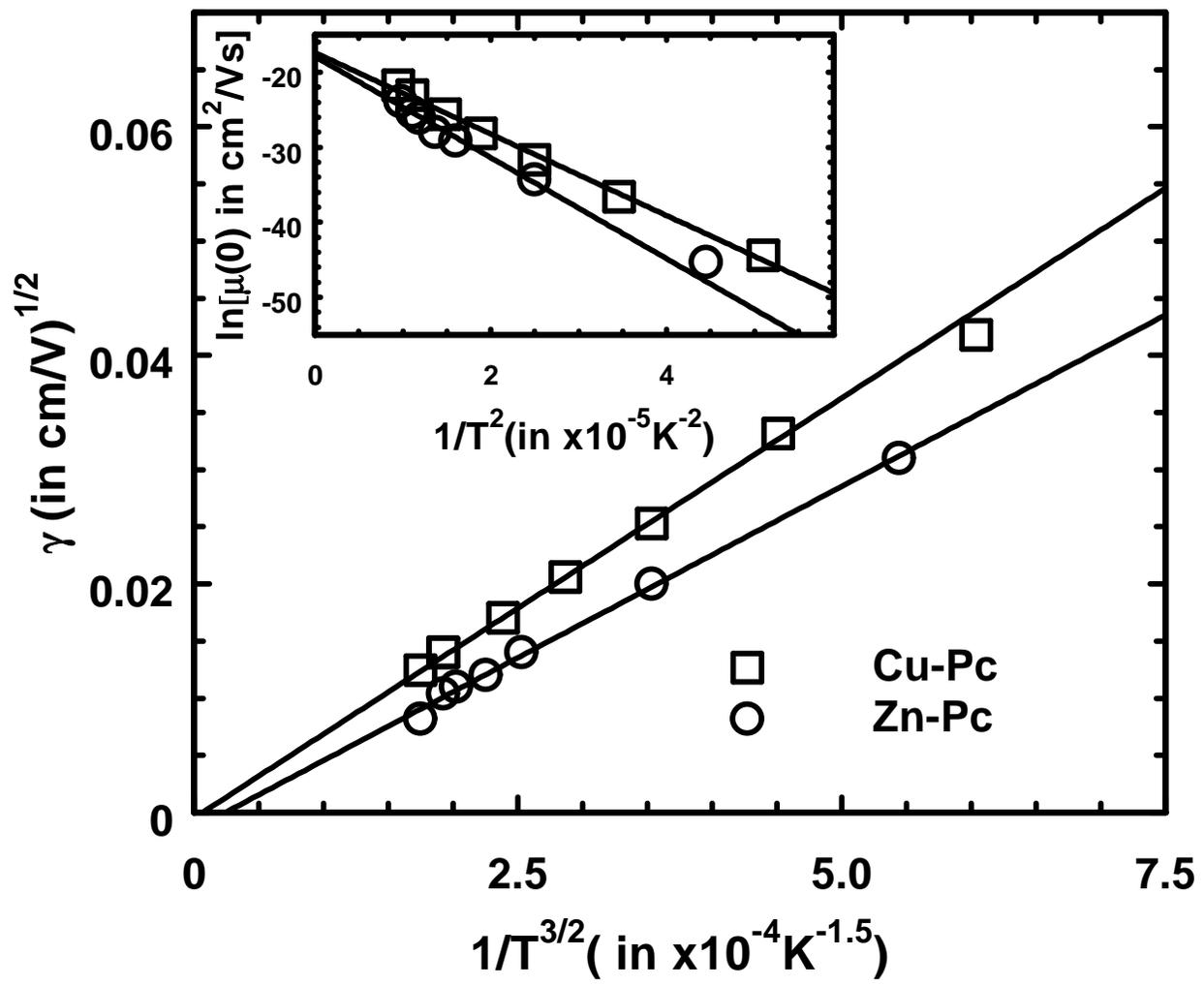
\end{document}